\documentclass{article}
\usepackage{geometry}
\usepackage[utf8]{inputenc}
\usepackage{color}
\usepackage{amsmath}
\usepackage{threeparttable}
\usepackage{graphicx}
\usepackage{float}
\usepackage{subfigure}
\usepackage[backref]{hyperref} 

\title{\textbf{Pricing Exchange Option Based on Copulas by MCMC Algorithm}}
\author{Wen Su}
\date{}

\begin{document}

\maketitle

\begin{abstract}
This paper focus on pricing exchange option based on copulas by MCMC algorithm. Initially, we introduce the methodologies concerned about risk-netural pricing, copulas and MCMC algorithm. After the basic knowledge, we compare the option prices given by different models, the results show except Gumbel copula, the other model provide similar estimation.
\end{abstract}


\section{Risk-Netural Pricing with C.D.F.}
A call option price can be expressed as an expectation (conditional expectation) under risk-netural measure $Q$:
\begin{equation}
c\left( t,S,T,K \right) =e^{-r\left( T-t \right)}E^Q\left[ \left. \left( S_T-K \right) ^+ \right|\mathcal{F} _t \right],
\end{equation}
this expectation can be simplified by using the conditional terminal c.d.f. of the underlying asset $S_T$ denoted by $F_t(\cdot)$, since a non-negative random variable $X$ enjoys a property
\begin{equation}
E\left( X \right) =\int_0^{+\infty}{P\left( X>x \right) dx}=\int_0^{+\infty}{\left[ 1-F_X\left( x \right) \right] dx}.
\end{equation}
So the call option price can be expressed as
\begin{equation}
\begin{aligned}
	c\left( t,S,T,K \right) &=e^{-r\left( T-t \right)}\int_0^{+\infty}{Q_t\left\{ \left( S_T-K \right) ^+>x \right\} dx}\\
	&=e^{-r\left( T-t \right)}\int_K^{+\infty}{Q_t\left( S_T>x \right) dx}\\
	&=e^{-r\left( T-t \right)}\int_K^{+\infty}{\left[ 1-F_t\left( x \right) \right] dx}.\\
\end{aligned}
\end{equation}

Thus in other words, pricing option is equal to finding $F_t(x)$. We differentiate (3) w.r.t. $K$ then we have
\begin{equation}
\frac{\partial c\left( t,S,T,K \right)}{\partial K}=e^{-r\left( T-t \right)}\left[ 1-F_t\left( K \right) \right],
\end{equation}
which may be helpful for finding $F_t(x)$.

\section{Copulas}
The copula developed by Sklar (1959) is a useful tool for handling multivariate distributions with given univariate marginals. Formally, a copula $C$ is a distribution function, defined on the unit cube $[0, 1]^d$, with uniform one-dimensional marginals. For example, a random vector $X=(X_1,\cdots,X_d)^T$ has joint c.d.f. $F(x_1,\cdots,x_d)$ and marginals $F_1,\cdots,F_d$, then copula is defined as
\begin{equation}
\begin{aligned}
C\left( u_1,\cdots ,u_d \right) &=P\left( F_1\left( X_1 \right) \le u_1,\cdots ,F_d\left( X_d \right) \le u_d \right) 
\\
&=F\left( F_{1}^{-}\left( u_1 \right) ,\cdots ,F_{d}^{-}\left( u_d \right) \right) ,
\end{aligned}
\end{equation}
where $u_i \in [0,1]$, $i=1,\cdots,d$. Generally, copula is considered to be the joint c.d.f. of $(F_1(X_1),\cdots,F_d(X_d))$. Thus, for multivariate distributions with continuous marginals, the univariate marginals and multivariate dependence structure can be separated, and the dependence structure can be represented by a copula. In some conditions, the survival copula is more effective. The survival copula is defined as
\begin{equation}
\bar{C}\left( u_1,\cdots ,u_d \right) =P\left( F_1\left( X_1 \right) >1-u_1,\cdots ,F_d\left( X_d \right) >1-u_d \right),
\end{equation}
where $u_i \in [0,1]$, $i=1,\cdots,d$. In real problems, $2$-d copulas are applied most often, so the following part will introduce some common $2$-d copulas.
\begin{enumerate}
\item Independent Copula:
$$
\Pi \left( u,v \right) = uv.
$$
\item Comonotonicity Copula:
$$
M \left( u,v \right) = \min\{u,v\}.
$$
\item Countermonotonicity Copula:
$$
m \left( u,v \right) = \max\{u+v-1,0\}.
$$
\item Frechet Copula:
$$
C^{Fre}\left( u,v \right) =\alpha M\left( u,v \right) +\beta \Pi \left( u,v \right) +\gamma m\left( u,v \right) .
$$
\item Gumbel Copula:
$$
C^{Gum}_{\theta}\left( u,v \right) =\exp \left\{ -\left[ \left( -\ln u \right) ^{\theta}+\left( -\ln v \right) ^{\theta} \right] ^{\frac{1}{\theta}} \right\}.
$$
\item Clayton Copula:
$$
C^{Cla}_{\theta}\left( u,v \right) =\left( u^{-\theta}+v^{-\theta}-1 \right) ^{-\frac{1}{\theta}}.
$$
\item Frank Copula:
$$
C^{Fra}\left( u,v \right) =-\frac{1}{\theta}\ln \left( 1+\frac{\left( e^{-\theta u}-1 \right) \left( e^{-\theta v}-1 \right)}{e^{-\theta}-1} \right) .
$$
\item Gaussian Copula: $(X,Y)\sim N(\mu_1,\mu_2;\sigma_1^2,\sigma_2^2;\rho)$, copula for it is
$$
C^{Gau}_{\rho}\left( u,v \right) =\frac{1}{2\pi \sqrt{1-\rho ^2}}\int_{-\infty}^{N^{-1}\left( u \right)}{\int_{-\infty}^{N^{-1}\left( v \right)}{e^{-\frac{x^2+y^2}{2\left( 1-\rho ^2 \right)}}dxdy}}.
$$
\end{enumerate}

\section{MCMC Algorithm}
Markov Chain Monte Carlo (MCMC) is a simulation method which applies Markov Chain into Monte Carlo procedure. In Monte Carlo simulation, if we want to calculate expectation $E[f(X)]$, we will try to produce random sample $x_1,x_2,\cdots,x_n$ from population $X$, usually $n$ is large, and then the law of large number ensures
\begin{equation}
\frac{1}{n}\sum_{i=1}^n{f\left( x_i \right)}\xrightarrow{P}E\left[ f\left( X \right) \right] .
\end{equation}

However, random sample from $X$ sometimes may be not easy to produced. But if a homogeneous Markov Chain sequence $x_1,\cdots,x_n,\cdots$ (with some technical conditions) which converges to $X$ can be produced, then we have
\begin{equation}
\frac{1}{n}\sum_{i=1}^n{f\left( x_i \right)}\xrightarrow{P}E\left[ f\left( X \right) \right].
\end{equation}
However, since currently $x_1,\cdots,x_n$ are not i.i.d., so it depends on $x_{t-1}$ to produce $x_t$, so the LHS of (8) may depend on the initial point $x_1$ (or $x_0$). So usually the estimate is given by 
$$\frac{1}{n}\sum_{i=m+1}^{m+n}{f\left( x_i \right)}.$$

Gibbs Sampler is one of the best-known MCMC methods to produce a Markov Chain sequence, it takes two steps to get $(x_{t+1},y_{t+1})$ from $(x_{t},y_{t})$:
\begin{enumerate}
    \item produce $x_{t+1}$ from $X \mid Y=y_t$;
    \item produce $y_{t+1}$ from $Y \mid X=x_{t+1}$.
\end{enumerate}
Continuing in this fashion, the Gibbs Sampler generates a sequence of random variables $\{(x_k,y_k),k=1,2,\cdots,n\}$ which converges to $(X,Y)$.
\section{Pricing Based on GBM}
An exchange option $V(t,S_1,S_2,T)$ is an European derivative contract with the terminal payoff
\begin{equation}
V\left( T,S_1,S_2,T \right) =\max \left\{ S_1\left( T \right) -S_2\left( T \right) ,0 \right\}.
\end{equation}

In so-called Black-Scholes economics, the underlying assets dynamics are given by GBM, i.e.,
\begin{equation}
\begin{aligned}
&\frac{dS_1\left( t \right)}{S_1\left( t \right)}=\mu _1dt+\sigma _1dB_1\left( t \right) ,\\
&\frac{dS_2\left( t \right)}{S_2\left( t \right)}=\mu _2dt+\sigma _2dB_2\left( t \right) ,\\
\end{aligned}
\end{equation}
besides, $B_1(t)$ and $B_2(t)$ are often not independent, $dB_1\left( t \right) dB_2\left( t \right) =\rho dt.$

There is an unique risk-netural measure $Q$ which ensures
\begin{equation}
\begin{aligned}
&\frac{dS_1\left( t \right)}{S_1\left( t \right)}=rdt+\sigma _1dB^Q_1\left( t \right) ,\\
&\frac{dS_2\left( t \right)}{S_2\left( t \right)}=rdt+\sigma _2dB^Q_2\left( t \right) ,\\
\end{aligned}
\end{equation}
and of course, $dB^Q_1\left( t \right) dB^Q_2\left( t \right) =\rho dt.$ Actually, we can decompose $dB_2(t)$ as $dB_2(t) = \rho dB_1(t) + \sqrt{1-\rho^2} dW(t)$ where $W(t)$ is a BM uncorrelated with $B_1(t)$ and then we can easily construct the Radon-Nikodym derivative.

Under risk-netural measure, the option price can be expressed as a conditional expectation
\begin{equation}
V\left( t,S_1,S_2,T \right) =e^{-r\left( T-t \right)}E^Q\left[ \left. \max \left\{ S_1\left( T \right) -S_2\left( T \right) ,0 \right\} \right|\mathcal{F} _t \right] ,
\end{equation}

Note that $\frac{e^{-rt}S_2\left( t \right)}{S_2(0)} =e^{\sigma _2B_{2}^{Q}\left( t \right) -\frac{\sigma _{2}^{2}t}{2}}$ can be used as a Radon-Nikodym derivative $Z_t=\left. \frac{d\nu}{dQ} \right|_{\mathcal{F} _t}$, which makes $B_2^{\nu}(t) = -\sigma_2 t + B_2^Q(t)$ a BM in $\nu$. So we have
\begin{equation}
\begin{aligned}
	V\left( t,S_1,S_2,T \right) &=e^{-r\left( T-t \right)}E^Q\left[ \left. \max \left\{ S_1\left( T \right) -S_2\left( T \right) ,0 \right\} \right|\mathcal{F} _t \right]\\
	&=e^{-r\left( T-t \right)}E^Q\left[ \left. S_2\left( T \right) \max \left\{ \frac{S_1\left( T \right)}{S_2\left( T \right)}-1,0 \right\} \right|\mathcal{F} _t \right]\\
	&=e^{rt}S_2\left( 0 \right) E^Q\left[ \left. Z_T\max \left\{ \frac{S_1\left( T \right)}{S_2\left( T \right)}-1,0 \right\} \right|\mathcal{F} _t \right]\\
	&=e^{rt}S_2\left( 0 \right) Z_tE^{\nu}\left[ \left. \max \left\{ \frac{S_1\left( T \right)}{S_2\left( T \right)}-1,0 \right\} \right|\mathcal{F} _t \right]\\
	&=S_2\left( t \right) E^{\nu}\left[ \left. \max \left\{ \frac{S_1\left( T \right)}{S_2\left( T \right)}-1,0 \right\} \right|\mathcal{F} _t \right] .\\
\end{aligned}
\end{equation}

By It\^{o}'s lemma, we have
\begin{equation}
\begin{aligned}
\frac{d\left( \frac{S_1}{S_2} \right)}{\frac{S_1}{S_2}}&=\left( -\rho \sigma _1\sigma _2dt+\sigma _1dB_{1}^{Q} \right) +\left( \sigma _2dt-\sigma _2dB_{2}^{Q} \right) 
\\
&=\left( \rho \sigma _1dB_{2}^{\nu}+\sigma _1\sqrt{1-\rho ^2}dB_{3}^{\nu} \right) -\sigma _2dB_{2}^{\nu}
\\
&=\sigma _1\sqrt{1-\rho ^2}dB_{3}^{\nu}+\left( \rho \sigma _1-\sigma _2 \right) dB_{2}^{\nu}
\\
&=\sqrt{\sigma _{1}^{2}+\sigma _{2}^{2}-2\rho \sigma _1\sigma _2}dW^{\nu}.
\end{aligned}
\end{equation}
So the expectation can be calculated by Black-Schoels formula,
\begin{equation}
E^{\nu}\left[ \left. \max \left\{ \frac{S_1\left( T \right)}{S_2\left( T \right)}-1,0 \right\} \right|\mathcal{F} _t \right] =\frac{S_1\left( t \right)}{S_2\left( t \right)}N\left( d_1 \right) -N\left( d_2 \right) ,
\end{equation}
where $d_1$ and $d_2$ are given by
$$
d_1=\frac{\ln \frac{S_1\left( t \right)}{S_2\left( t \right)}+\frac{\sigma _{1}^{2}+\sigma _{2}^{2}-2\rho \sigma _1\sigma _2}{2}\left( T-t \right)}{\sqrt{\left(\sigma _{1}^{2}+\sigma _{2}^{2}-2\rho \sigma _1\sigma _2\right)\left( T-t \right)}},\quad d_2=d_1-\sqrt{\left(\sigma _{1}^{2}+\sigma _{2}^{2}-2\rho \sigma _1\sigma _2\right)\left( T-t \right)} .
$$

Finally, we have the exchange option price
\begin{equation}
V\left( t,S_1,S_2,T \right) = S_1\left( t \right) N\left( d_1 \right) -S_2\left( t \right) N\left( d_2 \right) .
\end{equation}

\section{Pricing Based on Copulas}
Actually, the underlying assets dynamics may be not given by GBM. Learnt from section 1, we have
\begin{equation}
\begin{aligned}
	V\left( t,S_1,S_2,T \right) &=e^{-r\left( T-t \right)}E^Q\left[ \left. \left( S_1\left( T \right) -S_2\left( T \right) \right) ^+ \right|\mathcal{F} _t \right]\\
	&=e^{-r\left( T-t \right)}E^Q\left[ \left. S_1\left( T \right) -\min \left\{ S_1\left( T \right) ,S_2\left( T \right) \right\} \right|\mathcal{F} _t \right]\\
	&=S_1\left( t \right) -e^{-r\left( T-t \right)}E^Q\left[ \left. \min \left\{ S_1\left( T \right) ,S_2\left( T \right) \right\} \right|\mathcal{F} _t \right]\\
	&=S_1\left( t \right) -e^{-r\left( T-t \right)}\int_0^{+\infty}{Q_t\left( \min \left\{ S_1\left( T \right) ,S_2\left( T \right) \right\} >x \right) dx}\\
	&=S_1\left( t \right) -e^{-r\left( T-t \right)}\int_0^{+\infty}{\bar{C}_t\left( Q_t\left( S_1\left( T \right) >x \right) ,Q_t\left( S_2\left( T \right) >x \right) \right) dx}.\\
\end{aligned}
\end{equation}

However, survival copula for the asset prices may be not easy to find, let
\begin{equation}
S_1(T)=S_1(t)e^{X_1}\quad,S_2(T)=S_2(t)e^{X_2},
\end{equation}
then we can focus only on the copula of rate of return $X_1$ and $X_2$, we have
\begin{equation}
\begin{aligned}
	\int_0^{+\infty}{Q_t\left( \min \left\{ S_1\left( T \right) ,S_2\left( T \right) \right\} >z \right) dz}&=\int_0^{+\infty}{Q_t\left( X>\ln \frac{z}{S_1\left( t \right)},Y>\ln \frac{z}{S_2\left( t \right)} \right) dz}\\
	&=\int_0^{+\infty}{\bar{C}_t\left( 1-F_1\left( \ln \frac{z}{S_1\left( t \right)} \right) ,1-F_2\left( \ln \frac{z}{S_2\left( t \right)} \right) \right) dz}\\
	&=\int_0^{+\infty}{\left[ 1- u -v +C_t\left( u,v \right) \right] dz},\\
\end{aligned}
\end{equation}
where $F_1$, $F_2$ are the conditional c.d.f. for $X_1$ and $X_2$, respectively, $u_1=u_1(z)=F_1(\ln \frac{z}{S_1(t)})$, $v=v(z)=F_2(\ln \frac{z}{S_2(t)})$, and note that in $2$-d situation, survival copula can be expressed as $\bar{C}(1-u,1-v)=1-u-v+C(u,v)$.

From (19) we clearly see why pricing exchange option is linked to copula. However, in order to price exchange option, besides type of copula, we should also know the of marginal c.d.f. of $X$ and $Y$. Generally, marginal normal may be a common choice but remember the joint distribution may be not normal.

In order to calculate $E^Q\left[ \left. \left( S_1\left( T \right) -S_2\left( T \right) \right) ^+ \right|\mathcal{F} _t \right]$, consider $X_1\sim N(r(T-t),\sigma_1^2(T-t))$, $X_2\sim N(r(T-t),\sigma_2^2(T-t))$, $U_1\sim U(0,1)$, $U_2\sim U(0,1)$ and $(U_1,U_2)\sim C(u_1,u_2)$, we divide the procedure into following steps: 
\begin{enumerate}
    \item Draw $u_1^{(1)} \sim U(0,1)$;
    \item Draw $u_2^{(1)} \sim U_2 \mid U_1=u_1^{(1)}$;
    \item Draw $u_1^{(2)} \sim U_1 \mid U_2=u_2^{(1)}$;
    \item Draw $u_2^{(2)} \sim U_2 \mid U_1=u_1^{(2)}$;
    \item Continuing in this fashion, the Gibbs Sampler generates a sequence of random variables $$\{(u_1^{(k)},u_2^{(k)}),k=1,2,\cdots,n+m\}$$ which converges to $(U_1,U_2)$;
    \item Let $x_1^{(k)}= r(T-t) + \sigma_1 \sqrt{T-t} N^{-1}(u_1^{k)})$, $x_2^{(k)}= r(T-t) + \sigma_2 \sqrt{T-t} N^{-1}(u_2^{(k)})$, $k=1,2,\cdots,n+m$.
\end{enumerate}
Then the MCMC estimate of $E^Q\left[ \left. \left( S_1\left( T \right) -S_2\left( T \right) \right) ^+ \right|\mathcal{F} _t \right]$ is given by
$$
\frac{1}{n}\sum_{k=m+1}^{n+m}{\left( S_1\left( t \right) e^{X_1}-S_2\left( t \right) e^{X_2} \right) ^+}.
$$

In the above procedure, $r$ is the risk-free rate, $\sigma_1$, $\sigma_2$ are the volatilities for two assets. Besides, the parameters in the copula should be estimated, too. Actually, if we want to obtain the MLE of all parameters from data $(x_{11},x{21})$, $(x_{12},x_{22})$, $\cdots$, $(x_{1n},x_{2n})$, we should maximize the log likelihood function
$$
\sum_{k=1}^n{\ln c\left( F_1\left( x_{1k} \right) ,F_2\left( x_{2k} \right) \right)}+\sum_{k=1}^n{\left( \ln f_1\left( x_{1k} \right) +\ln f_2\left( x_{2k} \right) \right)},
$$
where $c(u_1,u_2)$ is the p.d.f. of $C(u_1,u_2)$. It may be too tedious, we take it into two parts: we firstly maximize
$$
\sum_{k=1}^n{\left( \ln f_1\left( x_{1k} \right) +\ln f_2\left( x_{2k} \right) \right)},
$$
then apply the estimated parameters and maximize
$$
\sum_{k=1}^n{\ln c\left( F_1\left( x_{1k} \right) ,F_2\left( x_{2k} \right) \right)}.
$$
This two-part method is called inference for the margins (IFM), which reduced the complexity in computing a lot.

\section{Empirical Results}
We pay attention to two underlying assets in China stock market: 000001.SZ as $S_1$ and 600325.SH as $S_2$. Expire $T-t$ is chosen to be $\frac{1}{4}$ year. 2018.7 - 2020.6 is the in-sample period and 2020.7 - 2020.12 is the out-of-sample period. Table 1 shows the estimation results.

\begin{table}
\centering
\caption{Estimation Results}
\setlength{\tabcolsep}{15mm}{
\begin{tabular}{cc}
\hline
Model                & Paremeters           \\ \hline
Marginal             & $\mu _1=0.3548,\mu _2=0.1018,\sigma _1=0.2023,\sigma _2=0.1920$          \\
Frechet              & $\alpha =0.7720,\beta =0,\gamma =0.2280 $                \\
Gumbel               & $\theta =4.0962$               \\
Clayton              & $\theta =2.9400$                  \\
Frank                & $\theta =17.5472$                \\
Gaussian             & $\rho =0.5439$               \\ \hline
\end{tabular}}
\label{Table 1}
\end{table}

Then we apply the different methods to calculate the price of exchange option, i.e., to calculate the expectation $e^{r(T-t)}E^Q[(S_1(T)-S_2(T))^+\mid \mathcal{F}_t]$ in the out-of-sample period 2020.7 - 2020.12 and compare them. The results can be seen in the Table 2 and Figure 1. We can see the results of Gaussian copula and GBM are very similar, actually they are numerical solution and analytical solution for one model, respectively. When $S_1(t)$ is far larger than $S_2(t)$ which means the option will be exercised at a high probability, the $5$ methods give very close results. But when $S_1(t)$ is near $S_2(t)$, the results of Gumbel copula appear too high. 

\begin{table}
\centering
\caption{Exchange Option Price Under Different Models}
\setlength{\tabcolsep}{7mm}{
\begin{tabular}{cccccc}
\hline
           & Gumbel & Clayton & Frank  & Gaussian & GBM    \\ \hline
2020-07-01 & 0.1233 & 0.0676  & 0.0683 & 0.0579   & 0.0555 \\
2020-07-02 & 0.1226 & 0.0651  & 0.0667 & 0.0519   & 0.0514 \\
2020-07-03 & 0.3082 & 0.1891  & 0.1848 & 0.1867   & 0.1854 \\
\multicolumn{6}{c}{$\cdots \quad \cdots \quad \cdots \quad \cdots \quad \cdots$}  \\
2020-09-23 & 1.8895 & 1.8172  & 1.7997 & 1.7839   & 1.7658 \\
2020-09-24 & 1.6961 & 1.6188  & 1.5881 & 1.5864   & 1.5741 \\
2020-09-25 & 2.0049 & 1.9629  & 1.9763 & 1.9708   & 1.9380 \\ \hline
\end{tabular}}
\label{Table 2}
\end{table}

\begin{figure}[ht]
\centering
\includegraphics[scale=0.5]{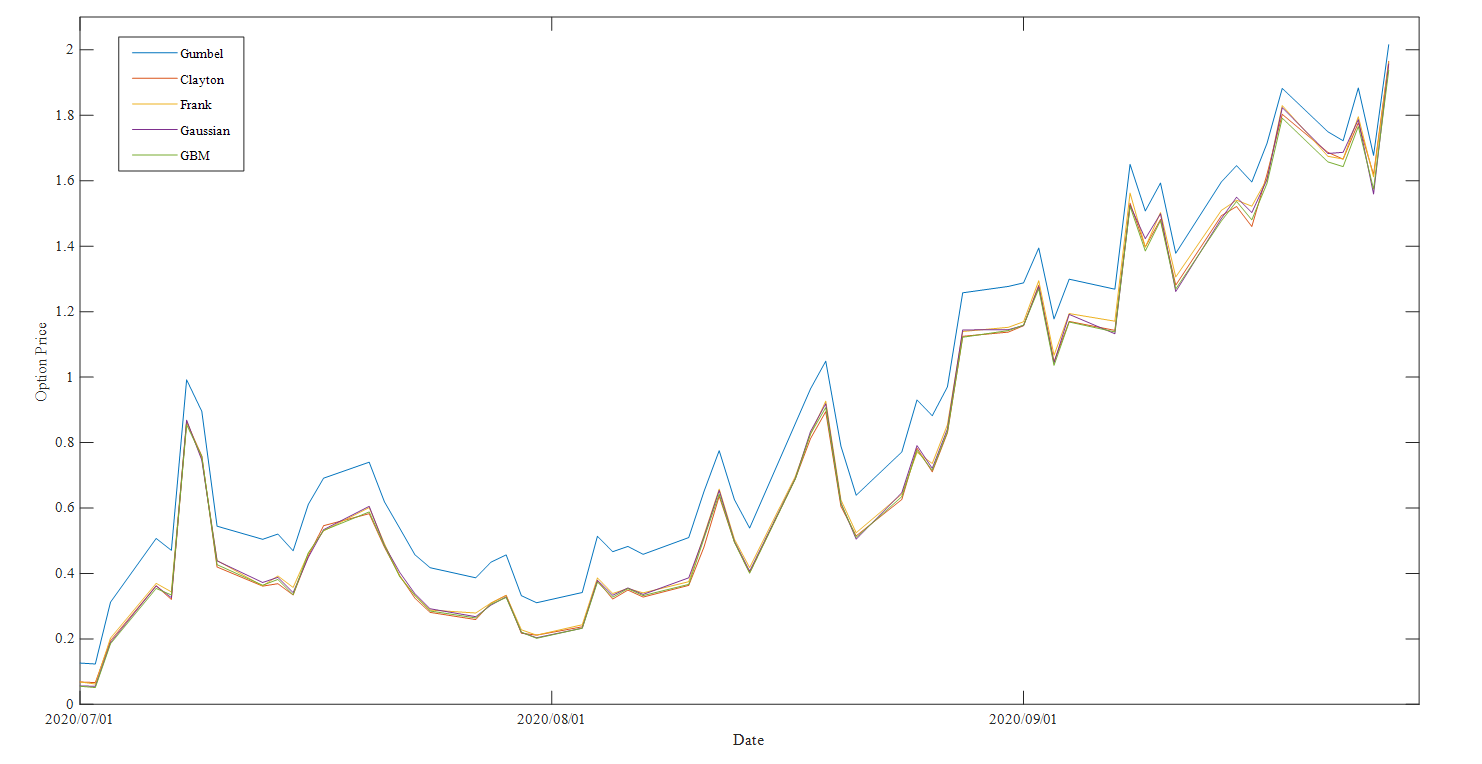}
\caption{\emph{Exchange Option Price Under Different Models}}
\label{Fig.1}
\end{figure}

\section{Conclusion}
Despite it seems copula model selection will exert a tremendous influence on pricing exchange option, the results show except Gumbel copula, the other models provide similar option prices. The facts we get in this research offer an insight that if we want to apply copula into pricing option (or c.d.f. method in section 1), any common type of copula may give similar results, so the estimation procedure (e.g: volatility, parameters in the copula) is still the most important thing. 

\section*{Acknowledgements}
Besides the mentioned papers, I also utilize some statements and mathematical operators from courseware of Prof. Xu Kai and Prof. Yang Jingping, Peking University as references.

\end{document}